\documentclass[]{jfm}

\def\ps@titlepage{%
  \def\@oddhead{}\def\@evenhead{}%
  \def\@oddfoot{}\def\@evenfoot{}%
  \def\sectionmark##1{}%
  \def\subsectionmark##1{}%
}

\let\absfooterflag\relax

\let\pagelimitfooter\relax
\makeatother

\usepackage{graphicx}
\usepackage{xcolor}
\usepackage{epstopdf}
\usepackage{array}
\usepackage{newtxtext}
\usepackage{newtxmath}
\usepackage{natbib}
\usepackage{hyperref}
\usepackage{soul}
\usepackage{placeins}
\usepackage{subcaption} 
\usepackage{tikz}
\hypersetup{
    colorlinks = true,
    urlcolor   = blue,
    citecolor  = black,
}

\title{
Bursting bubbles in Herschel–Bulkley fluids: 
dynamics and jetting transitions 
}

\author{A. H. Ghaemi\aff{1,2}, Z. Yang\aff{1}, A. Huang\aff{1}, V. Sanjay\aff{3}, J. Feng\aff{1}, C. R. Constante-Amores\aff{1}\corresp{\email{crconsta@illinois.edu}}}

\affiliation{
\aff{1}Department of Mechanical Science and Engineering, University of Illinois Urbana-Champaign, IL 61801, USA\\
\aff{2}Department of Flow Physics and Technology, Faculty of Aerospace Engineering, Delft University of Technology, 2629 HS Delft, The Netherlands\\
\aff{3}CoMPhy Lab, Department of Physics, Durham University,
Science Laboratories, South Road, Durham DH1 3LE, United Kingdom
}

\begin{document}
\maketitle

\begin{abstract}
When a bubble rises to a free surface, its bursting dynamics in Newtonian fluids are governed by the interplay between viscous, capillary, and gravitational forces. In this work, we extend this classical problem to Herschel–Bulkley fluids, elucidating the role of viscoplasticity and non-Newtonian rheology in bubble bursting. Using direct numerical simulations validated against experiments, we systematically explore the influence of the key governing dimensionless parameters, such as the Bond number, the Ohnesorge number, the shear-dependent behavior 
and the plastocapillary number, each varied over several orders of magnitude. Our results reveal that viscoplasticity strongly controls the evolution and interaction of capillary waves within the cavity formed upon bubble rupture. Shear-thinning and shear-thickening effects are significant only for moderate Ohnesorge numbers, while at large Ohnesorge values the free surface dynamics converge to a non-flat equilibrium shape once the internal stresses fall below the yield stress. 
These findings provide new insights into the coupled effects of viscosity, gravity, yield stress, and shear-dependent rheology in multiphase flows, with broad implications for natural and industrial processes involving gas-liquid interfaces.


\end{abstract}

\section{Introduction}

When a bubble at a liquid–gas interface bursts, it undergoes a sequence of events comprising film rupture, cavity collapse, capillary-wave focusing, and, in some cases, the formation of a Worthington jet.
This jet breaks up into small droplets, which could transport biological material, toxins, salts, surfactants or dissolved gases 
\citep{Woodcock_nature_1953,MacIntyre_jgr_1972,Poulain_prl_2018,bird_2023}.
Bubble bursting has drawn sustained scientific attention owing to its relevance across a wide range of natural phenomena and industrial operations. The fundamental mechanisms controlling the rupture of the liquid film and the ensuing ejection of droplets have been systematically explored through meticulous experiments \citep{Woodcock_nature_1953,Ghabache_pof_2014,Ghabache_prf_2016,Seon_epjst_2017}, high-fidelity numerical simulations \citep{Gordillo_jfm_2019,Singh_prf_2019,crca_bb}, and  theoretical analyses \citep{Zeff_nature_2000,Lai_prl_2018,Blanco-Rodriguez_jfm_2020}.

For Newtonian liquids, the collapse of the cavity is governed by a delicate interplay between inertia, viscosity, and surface tension. Upon rupture, capillary waves move downward the cavity  leading to the reverse of the interface, and resulting in the formation of a strong upward jet that may fragment into one or more droplets \citep{Zeff_nature_2000}. At low viscosity, the flow is largely inertial-capillary in nature, producing tall and slender jets that eject droplets at high speed. As viscosity increases, dissipation dampens the collapse, resulting in shorter, broader jets or complete suppression of ejection \citep{Lai_prl_2018}. A wealth of experimental, numerical, and theoretical studies have established these mechanisms, providing a coherent picture of how the bubble’s initial size, fluid properties, and ambient conditions dictate the jet height, velocity, and droplet formation dynamics \citep{Ghabache_prf_2016,Gordillo_jfm_2019,Blanco-Rodriguez_jfm_2020, berny2020role,jie_2025, deike2018dynamics,Ghabache_pof_2014}.

Beyond Newtonian fluids, bubble bursting in complex media such as viscoplastic, viscoelastic, or surfactant-laden liquids introduces additional rheological and interfacial effects that fundamentally alter the collapse dynamics \citep{sanjay2021bursting, Dixit_Oratis_Zinelis_Lohse_Sanjay_2025, crca_bb,feng_prl,cabalgante2025effect,Balasubramanian_Sanjay_Jalaal_Vinuesa_Tammisola_2024,barbhai2025effect}. 
In viscoplastic fluids, the presence of a finite yield stress modifies the force balance during cavity retraction by creating unyielded plug regions that restrict motion and dissipate energy \citep{Balmforth_Hewitt_2025}. The recent work of \citet{sanjay2021bursting} established the canonical picture of bubble bursting in viscoplastic media, showing that increasing the yield stress progressively suppresses jet formation and alters the final crater geometry. In the limiting Bingham case  in the absence of gravitational effects, their simulations revealed a transition from the classical inertial-capillary Worthington jet to complete jet suppression as the yield stresses are  increased. Subsequent analyses of the energy budget demonstrated that, unlike Newtonian fluids where most of the capillary energy converts to kinetic energy, viscoplastic systems dissipate a significant fraction of this energy in yielding the material, leading to residual crater shapes with finite curvature.
In viscoelastic liquids, the collapse is further influenced by elastic stresses that develop as polymer chains are stretched during cavity retraction, often delaying or arresting the collapse and giving rise to broader or slower jets \citep{Dixit_Oratis_Zinelis_Lohse_Sanjay_2025,cabalgante2025effect}. 
Similarly, surfactant-covered interfaces introduce Marangoni stresses that oppose local variations in surface tension, thereby dampening capillary focusing, suppressing droplet ejection, and inducing asymmetries in the jet \citep{crca_bb,pico2024surfactant,Pierre}. Together, these non-Newtonian and interfacial effects govern how surface energy is redistributed among viscous dissipation, elastic storage, and interfacial stresses, highlighting the need for a unified understanding of bubble bursting across complex fluids.

As revealed by the foregoing review, although bubble bursting in Newtonian liquids and certain non-Newtonian cases has been examined, the dynamics in yield-stress fluids remain far less understood, particularly when the material also exhibits shear-dependent viscosity. In Herschel–Bulkley media, shear-thinning and shear-thickening behavior profoundly modify the local resistance to flow during cavity collapse and jet formation. This interplay between, viscosity, capillarity,  yield stress and nonlinear rheology determines whether the cavity can accelerate into a jet or remains arrested in a partially yielded state. In this work, we investigate these mechanisms using fully resolved numerical simulations,  revealing how variations in the degree of shear-thinning or thickening reorganize capillary-wave dynamics, and the final surface morphology. The remainder of this paper is organized as follows: Section~\ref{Numerical} outlines the numerical approach and validation, Section~\ref{Results} presents the main findings, and concluding remarks are given in Section~\ref{Con}.

\section{Formulation, experimental and  numerical methodology \label{Numerical}}


\subsection{Governing Equations}

The numerical simulations were performed by solving the two-phase Navier-Stokes equations. Distances, times and pressures are non-dimensionalized using characteristic values, $R$, $\sqrt{\rho_l R_0^3/\sigma}$ and $\sigma/R$: hence, velocities are non-dimensionalized using the capillary velocity $U_\gamma=\sqrt{\rho R^3/\sigma}$. Asymmetric simulation have been performed using the open source code Basilisk \citep{popinet2009accurate, basilisk}. As a result of this scaling, the dimensionless equations read  

\begin{equation}
\nabla \cdot \mathbf{u} = 0, \quad 
\frac{\partial \mathbf{u}}{\partial t} + \mathbf{u} \cdot \nabla \mathbf{u} 
= -\nabla p + \nabla \cdot \boldsymbol{\tau} - Bo\, \mathbf{e}_z + \kappa \mathbf{n} \delta_s,
\label{eq:ns_dimless}
\end{equation}
here, $\mathbf{u}$ is the velocity field, $p$ is the pressure, $\boldsymbol{\tau}$ is the deviatoric stress tensor, $\kappa$ is the interface curvature, $\mathbf{n}$ is the unit normal to the interface, and $\delta_s$ is the surface delta function.
The gas phase is modeled as a Newtonian fluid with constant viscosity and density, while the bulk phase is modeled as a Herschel–Bulkley fluid characterized by a yield stress $\tau_y$, consistency $K$, and flow index $n$. Here, the stress tensor for such a fluid is given by:
\begin{equation}
\boldsymbol{\tau} = 2 \left[ \frac{\mathcal{J}}{2\|\mathbf{\mathcal{D}}\| + \varepsilon} \mathbf{I} + Oh_K (2\|\mathbf{\mathcal{D}}\| + \varepsilon)^{n - 1} \right] \mathbf{\mathcal{D}}
\end{equation}
where $\|\mathbf{\mathcal{D}}\|$ is the second invariant of the deformation rate tensor, and $\varepsilon$ a regularization parameter (discussed below).   The three dimensionless numbers governing the dynamics are:

\begin{equation}
\mathcal{J} = \frac{\tau_y R_0}{\sigma}, \quad 
Oh_K= \frac{K {\dot{\gamma}}^{n-1}} {\sqrt{\rho_l \ \sigma R_0}}, \quad 
Bo = \frac{\rho_l g R_0^2}{\sigma}
\label{eq:dimless_groups}
\end{equation}

The Ohnesorge number $Oh_K$ compares inertial–capillary to inertial–viscous timescales. Eliminating the explicit dependence on the shear rate, $\dot{\gamma}$, by choosing the characteristic
capillary--inertial rate 
$\dot{\gamma}_c \sim \sqrt{\sigma/\rho_l R^{3}}$,
the generalized Ohnesorge number becomes
$Oh_K = K  \sigma^{\frac{n-2}{2}}  \rho_l^{-\frac{n}{2}}  R_0^{-\frac{3n-2}{2}}$. The plastocapillary number $\mathcal{J}$ quantifies the competition between yield stress and capillary stress. The Bond number $Bo$ measures the ratio of gravitational to capillary forces. These non-dimensional groups control the interface dynamics, jet formation, and arrest behavior observed during bubble bursting. Finally, the density and viscosity ratios, defined as $\rho_r = \rho_g / \rho_l$ and $\mu_r = \mu_g / \mu_l$, are fixed in this study at $10^{-3}$ and $2 \times 10^{-2}$, respectively.

\subsection{Numerical setup }

When a bubble rests underneath of a free surface, its initial  quasi-static equilibrium interfacial shape can be obtained by solving the Young-Laplace equation 
\citep{toba1959drop, princen1963shape, sanjay2022viscous, villermaux2022bubbles}, which depends solely on the Bond number.   In this work, we simulate a wide range of Bond numbers, from $Bo = 10^{-3}$ to $Bo = 1$, to cover both capillary- and gravity-dominated regimes, using the solution of the Young–Laplace equation to determine the appropriate initial bubble shape.
We note, however, that the Young–Laplace equation is derived for Newtonian liquids, whereas we consider HB fluids. In a viscoplastic medium, a bubble rises only if the buoyancy force exceeds the yield stress,
i.e., $Bo \sim  J$, which can produce complex, non-spherical shapes. In such cases, accurately specifying the initial condition for the bursting problem requires solving the full rise dynamics rather than relying on the quasi-static solution.

We also assume that the thin liquid cap ruptures instantaneously at $t = 0$, i.e., the bubble is initialized without the thin film. This cap-removal assumption is commonly adopted in bursting simulations and allows focusing purely on post-rupture dynamics \citep{deike2018dynamics, gordillo2019capillary, sanjay2021bursting,pico2024surfactant}). For small Bond numbers, the cap is extremely thin and its removal has negligible influence on the resulting dynamics (as shown in \citet{deike2018dynamics}). However, at large Bond numbers, the bubble cap can significantly influence the cavity collapse, affecting the formation and selection of dominant  capillary waves, and consequently, the jet velocity and the size of the ejected droplets. We note that the effect of the bubble cap is beyond the scope of this work but clearly deserves further investigation.

The computational domain is axisymmetric and spans $r, z \in [-5R, 5R]$. A large domain size is used to prevent reflections from the boundaries that could interfere with the collapse dynamics and jet formation, this choice is consistent with previous numerical studies
\citep{sanjay2021bursting,pico2024surfactant,gordillo2019capillary}. Symmetry is imposed at $r = 0$, a no-slip boundary condition is applied at the bottom wall, and outflow (Neumann) conditions are set at the top and right boundaries.   The computational domain is discretized using adaptive mesh refinement (AMR).  To construct the initial regime map shown in figure~\ref{fig:regime_map}, we use a refinement level of $10$,  such that the minimum mesh size is defined as $\Delta_{\min} = 10R_0/2^{10}$.  At this resolution, more than $800$ simulations were performed. 
For cases involving droplet ejection, and to more accurately redefine the  boundaries between different phenomenological behaviours, we employ higher levels of refinement,  up to a maximum level of $12$. This corresponds to a smallest cell size of  $\Delta_{\min} \approx 7.8 \times 10^{-3} R_0$, which provides approximately  $410$ cells across the bubble radius. Refinement is dynamically applied around the interface, velocity gradients, and curvature, while the far field is kept coarser \citep{sanjay2021bursting}. This resolution was verified to be sufficient for capturing cavity collapse and jet dynamics, as further refinement produced negligible changes in jet height and apex trajectory.

The Herschel--Bulkley constitutive law is implemented via a regularized formulation, in which the apparent viscosity is smoothed through an $\varepsilon$-dependent function to avoid singularities as $\| \mathbf{\mathcal{D}} \| \to 0$. As a result, the yield surface cannot be sharply resolved, since $\| \mathbf{\mathcal{D}} \|$ remains strictly positive for any finite $\varepsilon > 0$. We use a face implementation of the regularisation method, described in \cite{sanjay2025burstingbubble}. Nevertheless, regions where $\| \mathbf{\mathcal{D}} \|$ falls below a prescribed threshold can be interpreted as approximately unyielded, with an effective viscosity given by
\[
\mu_{\text{eff}} = 
\frac{\tau_y}{2\|\mathbf{\mathcal{D}}\|+\varepsilon}
+ K\,(2\|\mathbf{\mathcal{D}}\|+\varepsilon)^{\,n-1}.
\]
In all the simulations presented in the article, we have selected $\varepsilon = 10^{-2}$. We also note that this regularization approach has been successfully employed to study the pinch-off dynamics of viscoplastic threads, accurately recovering the limiting Newtonian behavior in the limiting case of $\mathcal{J}\rightarrow0$  \citep{yang2025pinch}.
See appendix for a detailed study of the dependency of $\varepsilon$ for a base case.

\subsection{Experimental setup}

\begin{figure}
    \centering
    \includegraphics[width=0.75\textwidth]{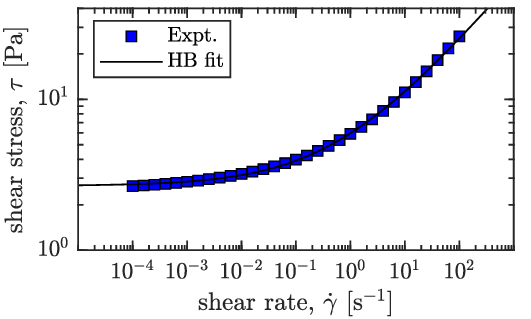}
        \caption{Rheological characterization of 0.1\% Carbopol 981 solution. The blue markers show the experimental steady state curve for the shear stress as a function of shear rate, and the solid curve shows the Herschel-Bulkley fitting.}
    \label{fig:rheol}
\end{figure}

To validate the simulation, we performed experiments of bubble bursting in a yield-stress fluid for comparison. A 0.1 wt\% neutralized solution of Carbopol (981, Lubrizol) was used as the yield stress fluid. The solution was prepared by dissolving the Carbopol in deionized water (resistivity of $18.2\ \rm{M\Omega}\ cm$, Smart2Pure 3 UV/UF, Thermo Fisher Scientific), followed by neutralization with 1 M NaOH (Emplura) solution to achieve a pH $=7$  \citep{nelson2017design}. Shear rheometry of the Carbopol solution was performed on an ARES-G2 rheometer using a 25 mm parallel-plate geometry, with an adhesive-backed 600 grit sandpaper (McMaster-Carr) on both bounding surfaces to mitigate wall slip \citep{ewoldt2014experimental,hossain2025critical}. Figure \ref{fig:rheol} shows the experimental data along with the Herschel-Bulkley model of $\tau = \tau_y+K\dot{\gamma}^n$, from which we obtained $\tau_y = 2.67~ \mathrm{Pa}, K = 3.29~\mathrm{Pa}\ \mathrm{s}^{n}$ and $n = 0.4187$ by fitting. Here $\tau, \tau_y, K, n$ and $\dot{\gamma}$ represent the shear stress, shear yield stress, consistency index, flow behavior index and shear rate, respectively. The bubble bursting experiments were performed in a container of $20\times 20\times 25\ \rm{mm}^3$. The bubble was generated by injecting air through a stainless-steel needle (inner diameter of 3.38 mm) using a syringe pump (11 Pico Plus Elite, Harvard Apparatus). After the bubble stabilized at the liquid surface, we punctured the bubble cap film with a clean, sharp stainless-steel needle and initiated the bubble rupture. The bubble rupture process was captured by a high-speed camera (Photron Mini AX100) at 6400 frames per second with a magnification of $\times1.0$. The images were post-processed with ImageJ and MATLAB to obtain the initial shape of the surface bubble and the meniscus.

\section{Results \label{Results}}

\begin{figure}
    \centering
    \begin{tabular}[t]{ccc}
        \begin{subfigure}{0.25\textwidth}
            \centering
            \includegraphics[width=\linewidth,trim={0 2.5cm 0 0},clip]{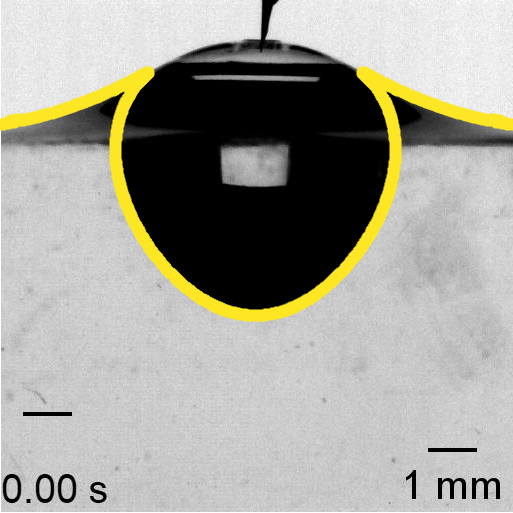} 
            \caption{}
        \end{subfigure} &
        \begin{subfigure}{0.25\textwidth}
            \centering
            \includegraphics[width=\linewidth,trim={0 2.5cm 0 0},clip]{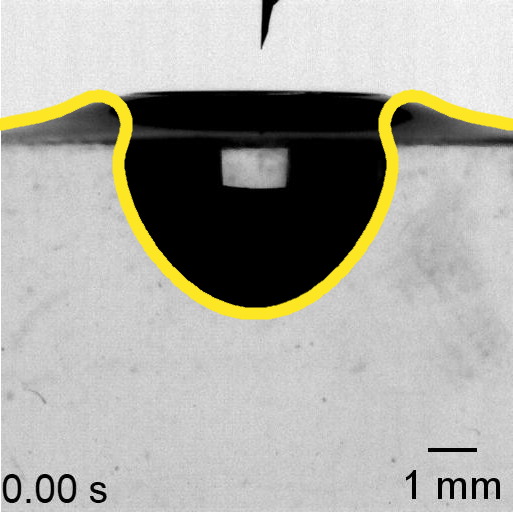}
            
            \caption{}
        \end{subfigure} &
        \begin{subfigure}{0.25\textwidth}
            \centering
            \includegraphics[width=\linewidth,trim={0 2.5cm 0 0},clip]{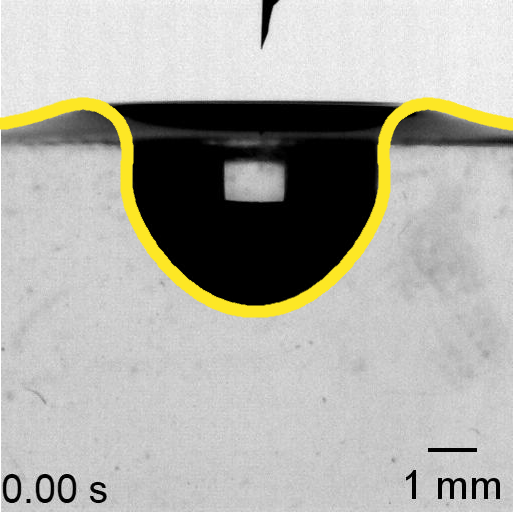}
            \caption{}
        \end{subfigure} \\
        \begin{subfigure}{0.25\textwidth}
            \centering
            \includegraphics[width=\linewidth,trim={0 2.5cm 0 0},clip]{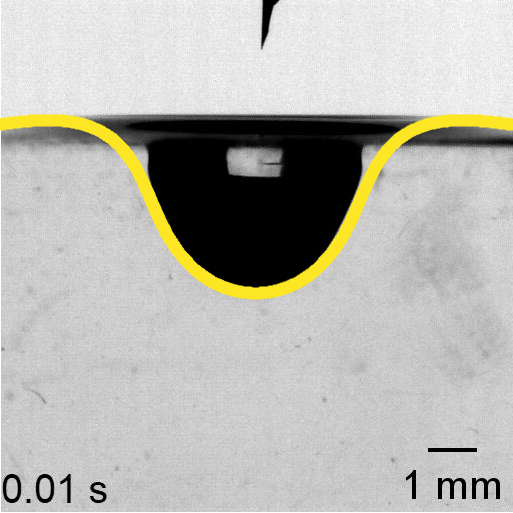}
            \caption{}
        \end{subfigure} &
        \begin{subfigure}{0.25\textwidth}
            \centering
            \includegraphics[width=\linewidth,trim={0 2.5cm 0 0},clip]{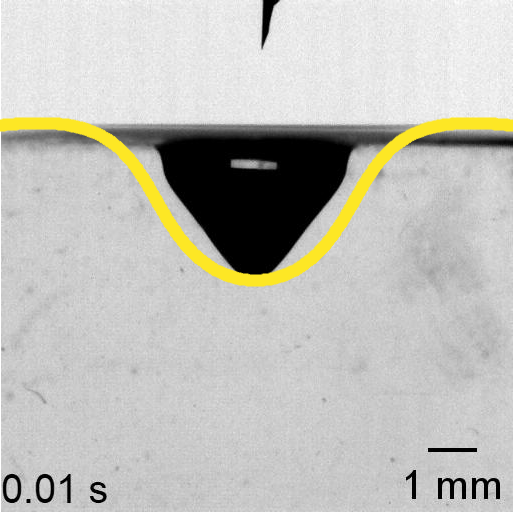}
            \caption{}
        \end{subfigure} &
        \begin{subfigure}{0.25\textwidth}
            \centering
            \includegraphics[width=\linewidth,trim={0 2.5cm 0 0},clip]{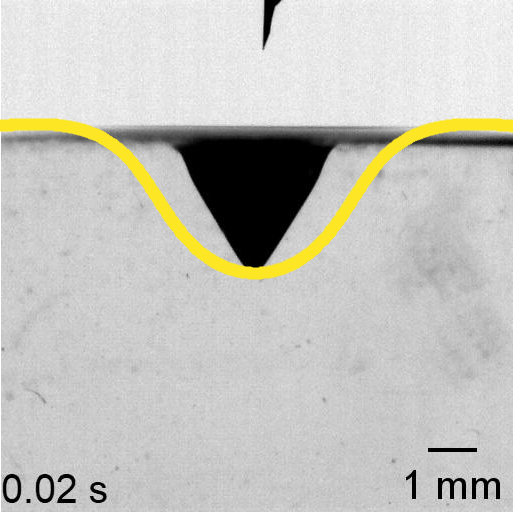}

            \caption{}
        \end{subfigure}
    \end{tabular}
    \caption{Temporal bursting bubble dynamics for a case with $n = 0.4187$, $J=0.1053$, $Bo=1.0990$ and $Oh_K=0.7015$ at (a) 0 ms, (b) 1.6 ms, (c) 2.8 ms, (d) 5.0 ms, (e) 7.5 ms, and (f) 21.6 ms. Numerical prediction of the air-liquid interface is overlapped on yellow in each experimental panel. The scale bar represents 1 mm.}
    \label{fig:validation_frames}
\end{figure}

We begin the Results section by comparing with the experimental case to validate our numerical code. The material properties of the Carbopol solution lead to dimensionless parameters of  $n = 0.4187$, $J=0.1053$, $Bo=1.0990$ and $Oh_K=0.7015$. In figure ~\ref{fig:validation_frames}, we 
show a one-to-one comparison   between experimental and DNS predictions. We display the liquid–air interfaces (yellow lines), which allow direct comparison between experiments and simulations. The initial bubble shape used for the simulations  is taken directly from the experiments (see figure \ref{fig:validation_frames}a).  At this set of parameters, the experiments show that the bursting lead to a
cavity collapses without reversal of the interface (i.e. no jet formation). This outcome can be attributed to a combination of large viscous effects, gravitational forces, shear-thinning behavior, and the viscoplastic nature of the fluid, as all four effects play a significant role. In the experiments, the post-collapse dynamics lead to the formation of  a pointed apex (see figure \ref{fig:validation_frames}f), which gradually retracts to form a nearly flat liquid–gas interface. While the DNS captures the early time evolution accurately, it does not reproduce the pointed interfacial shape observed in the experiments. Systematic variation of $\mathcal{J}$ and $n$ within the bounds of experimental uncertainty did not reproduce this pointy feature. We note that the long time discrepancies may be attributed to the weak viscoelastic character of Carbopol, whereas the present simulations assume a purely Herschel-Bulkley rheology \citep{hossain2025critical}. The formation of a pointed apex and its gradual relaxation are characteristic of an elastic rather than a purely viscoplastic response, suggesting that the weak elasticity of Carbopol may influence the late time interfacial dynamics.

Next, we perform a comprehensive parametric study of bubble bursting in viscoplastic media to elucidate how gravity, viscosity, shear-dependent rheology, and yield stress jointly govern the cavity collapse and jet formation. We perform over 800 simulations to explore the ranges of the four dimensionless parameters governing the dynamics:   $0.001 < Oh_K < 1$ (spanning inertia- to viscosity-dominated regimes), $0.01 < Bo < 1$ (from nearly spherical to gravity-deformed bubbles), $0.4 < n < 1.4$ (covering shear-thinning to shear-thickening behaviour), and $0 < \mathcal{J} < 0.7$ (from Newtonian to strongly viscoplastic). These ranges enable a systematic investigation of how each physical parameter modulates jet formation and droplet ejection.

\begin{figure}
\centering
\includegraphics[width=\linewidth]{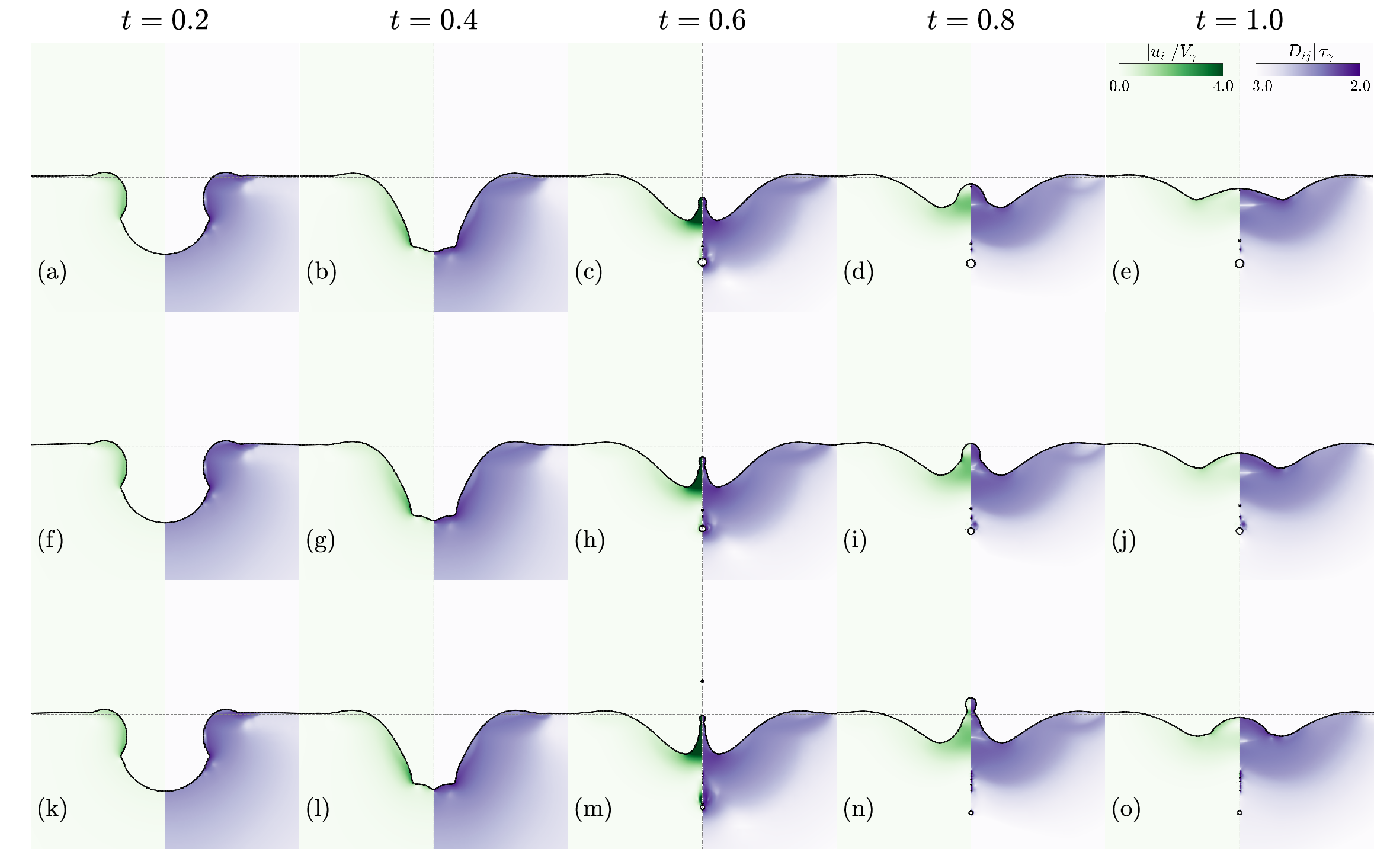}
\caption{Temporal evolution of bubble bursting in Herschel--Bulkley fluids for representative combinations of  parameters $n=0.8$, $Oh_K=0.01$, and $Bo=0.1$. The three rows correspond to different outcomes identified in the regime map of figure~4: (a–e) no jet with $\mathcal{J}=0.24$, (f–j) jet without droplet formation with $\mathcal{J}=0.22$, and (k–o) jet and droplet ejection with $\mathcal{J}=0.20$. For each case, panels show successive instants during the rupture process, with times nondimensionalised by the capillary timescale $\tau$. Each snapshot displays the velocity field (left) and the deformation–rate tensor modulus $\|\mathbf{\mathcal{D}}\|$ (right), together with the instantaneous interface position.}
\label{fig:temporal_evolution}
\end{figure}

\begin{figure}
    \centering
    \includegraphics[width=0.95\linewidth]{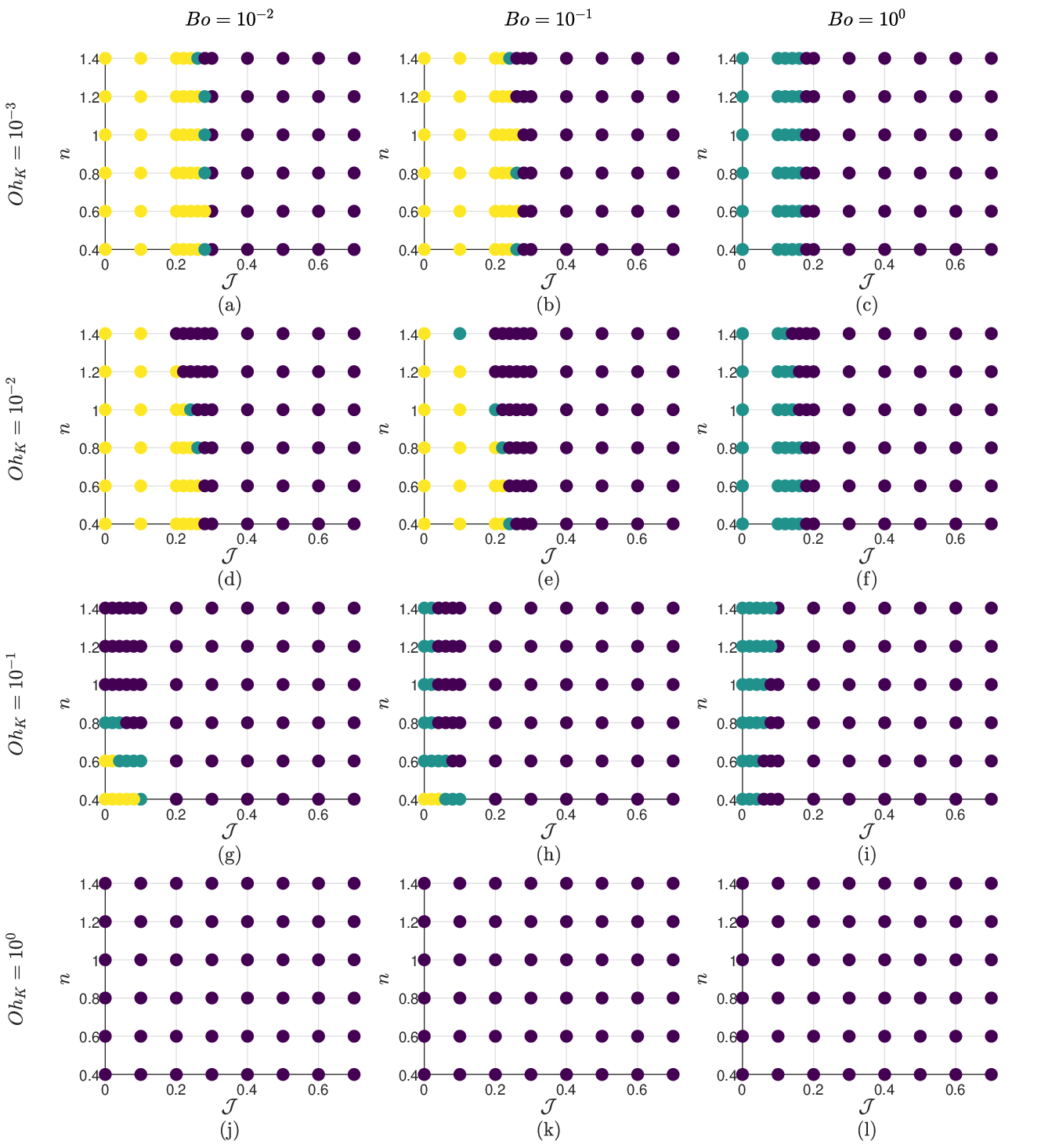}
    \caption{Regime map of bubble bursting in Herschel--Bulkley fluids in  the $(J,n, Oh_K, Bo)$ space. The horizontal axis represents the plastocapillary number $\mathcal{J}$ and the vertical axis the flow behavior index $n$. Each panel corresponds to a fixed pair of $Oh_K$ and $Bo$, as indicated. The brightest markers indicate cases with jetting and droplet formation, intermediate shading corresponds to jetting without droplet formation, and the darkest markers represent cases where neither a jet nor droplets are observed.}
    \label{fig:regime_map}
\end{figure}

\begin{figure}
\centering
\includegraphics[width=\linewidth]{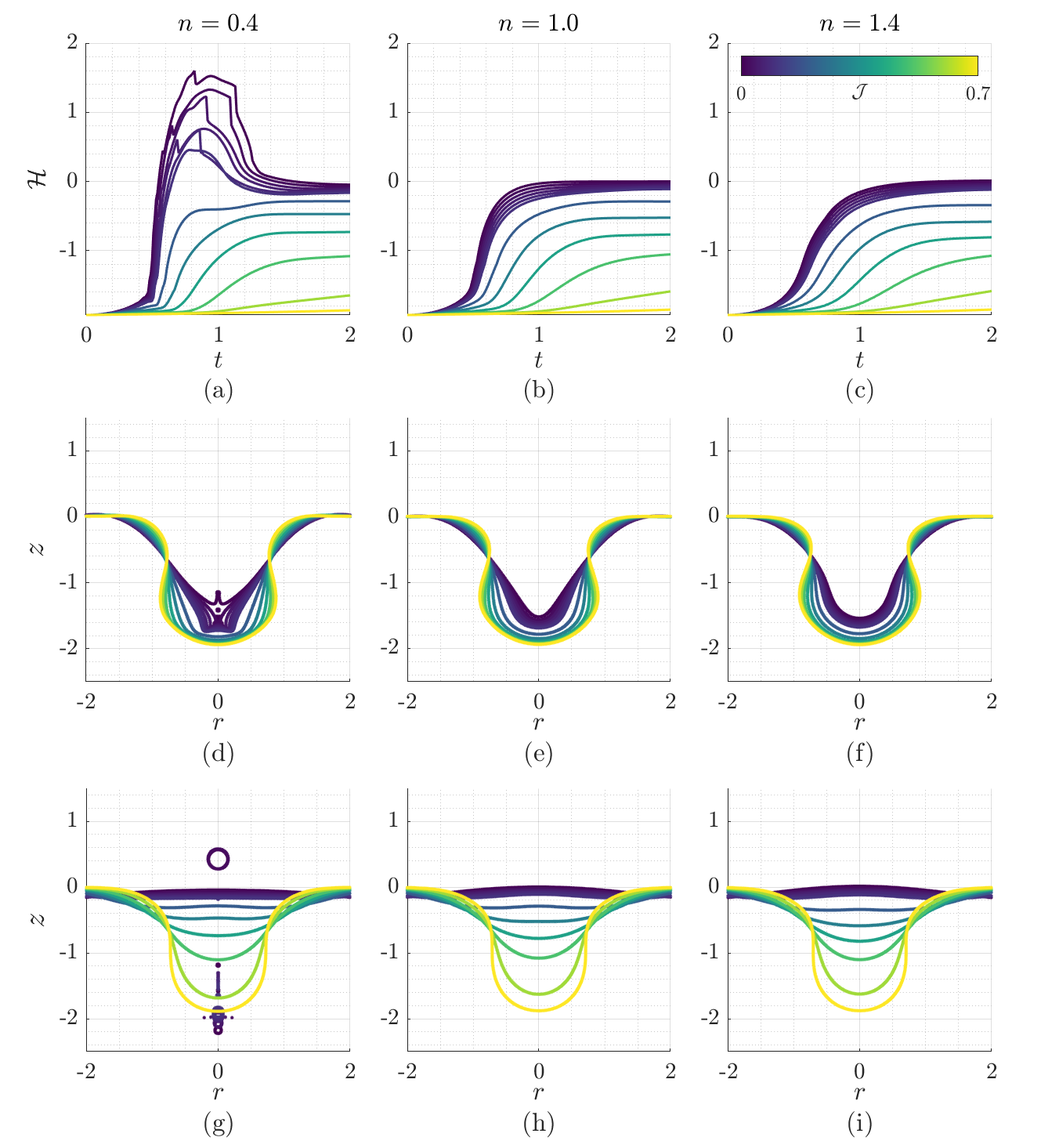}
\caption{Effect of $\mathcal{J}$ for $n=(0.4,1.0,1.4)$ with $Bo=0.01$ and $Oh_K=0.1$, corresponding to columns one-to-three, respectively. Panels (a–c) show the jet tip height $\mathcal{H}(t)$. Panels (d–f) and (g–i) display the interfacial shape at $t=0.5$ and $t=2.0$, respectively. The colour indicates the value of $\mathcal{J}$, as specified in the legend in panel (c).}
\label{fig:jet_height_shapes_J}
\end{figure}

Before discussing the regime maps as a function of the different dimensionless groups, figure~\ref{fig:temporal_evolution} shows three representative cases for $n = 0.8$, $Oh_K = 0.01$, and $Bo = 0.1$, which were selected to illustrate the progressive change in interfacial dynamics as 
$\mathcal{J}$ increases, serving as a reference for the regime maps discussed below. The rows correspond to plastocapillary numbers $\mathcal{J} = 0.24$, $0.22$, and $0.20$, while the columns show the spatial temporal evolution. 
For each snapshot, the left panels show the normalized velocity magnitude $|\mathbf{u}|/U_\gamma$, while the right panels display the normalized strain rate magnitude $|\mathbf{\mathcal{D}}_{ij}|\tau_\gamma$.
At  $\mathcal{J}=0.24$, the large yield stress  damps the interfacial motion, leading to  a no jet interfacial response, where the jet produced after cavity collapse is too weak to drive the surface above  the initial bulk level (see figure~\ref{fig:temporal_evolution}d).
At $\mathcal{J}=0.22$, the interface rises above the initial bulk level (see figure~\ref{fig:temporal_evolution}d), which we identify as the onset of jetting, although no droplet detaches from the tip.
Finally, at $\mathcal{J}=0.20$, the  yield stress is low enough for the  collapsing cavity to drive a strong fluid upward motion, culminating in the formation of a Worthington jet that ultimately pinches off to detach a droplet from its tip  (see figure~\ref{fig:temporal_evolution}m). This sequence highlights how small variations in $\mathcal{J}$ control the balance between capillary driving and yield stress resistance, leading to qualitatively distinct outcomes determined by the competition between capillarity, viscosity, and yield stress.

Figure~\ref{fig:regime_map} shows a phenomenological regime map in the $(\mathcal{J}, n, Oh_K, Bo)$ space, summarizing the outcomes of bubble bursting as based in figure ~\ref{fig:temporal_evolution}. Three distinct regimes are identified: no jet formation (i.e.. no upward motion), jet formation that rises above the initial bulk level without droplet ejection, and jet formation that culminates in droplet detachment.
At low $Bo$ (gravitational effects are weak leading to nearly spherical cavities) and  low $Oh$ (low viscous damping), it  leads  to a symmetrical collapse of the cavity.
The ensuing rapid inertial collapse favours the formation of a slender Worthington jet and, in some cases, droplet detachment (see figure \ref{fig:regime_map}a). As $Bo$  increases, 
gravity promotes bubble flattening and any elongation instead arises from yield-stress confinement during ascent
These deformations alter the collapse dynamics, thereby influencing both the strength of the ensuing jet and the likelihood of droplet detachment  (see figure \ref{fig:regime_map}c).
Similarly, for  $Oh_K=10^{-3}$, weak viscous damping allows for  stronger jetting and frequent  droplet formation, whereas at  $Oh_K=1$, strong viscous effects result in high energy dissipate rate and suppress jet formation altogether (see figure \ref{fig:regime_map}j-l). At intermediate $Oh_K=10^{-1}$, the full range of behaviors is recovered, spanning from no jet formation to jetting with  droplet detachment. Within each fixed combination of $Bo$ and $Oh_K$, decreasing either  $\mathcal{J}$ or $n$ shows that both weaker yield stress and stronger shear thinning behaviour promote jet ejection. The influence of $\mathcal{J}$ is particularly pronounced, as decreasing $\mathcal{J}$ reduces the the yield stress resistance, allowing capillary forces to drive the cavity collapse more efficiently and promote jet formation. In contrast, higher $\mathcal{J}$ values suppress interfacial motion, arresting the collapse before a jet can emerge. 
Variations in $n$ further modulate jetting, but their influence depends on $Oh_K$. At intermediate $Oh_K \sim 0.1$, shear-thinning fluids ($n < 1$) exhibit reduced effective viscosity at high strain rates, leading to weaker viscous dissipation and the formation of faster more elongated jets. In contrast, shear-thickening fluids ($n > 1$) increase local viscosity,  thereby damping the motion and inhibiting jet formation (see figure~\ref{fig:regime_map}g–i). At very low ($Oh_K \sim 0.001$) or very high ($Oh_K \gtrsim 1$) $Oh_K$, viscous effects are either negligible or dominant, rendering the dependence on $n$ comparatively weak. Together, these trends delineate the physical mechanisms that separate the regimes of no jetting, jetting without droplet formation, and droplet producing jets.

Given the breadth of the parameter space, next we focus on the temporal evolution of the jet height (e.g., measured as the interface location at the apex of symmetry), and the 
interface projections in  the $(r,z)$ plane at  two representative times, $t = 0.5$ and $t = 2$, as functions of the four dimensionless parameters. Examining the interface at these stages capture the competition between yield stress, viscosity, capillarity, and gravity that determines the cavity collapse intensity, jet formation, and final interface shape at long times.  This representation offers a concise overview of how variations in $\mathcal{J}$, $n$, $Oh_K$, and $Bo$ affect the collapse dynamics and resulting interface morphology.

\begin{figure}
\centering
\includegraphics[width=\linewidth]{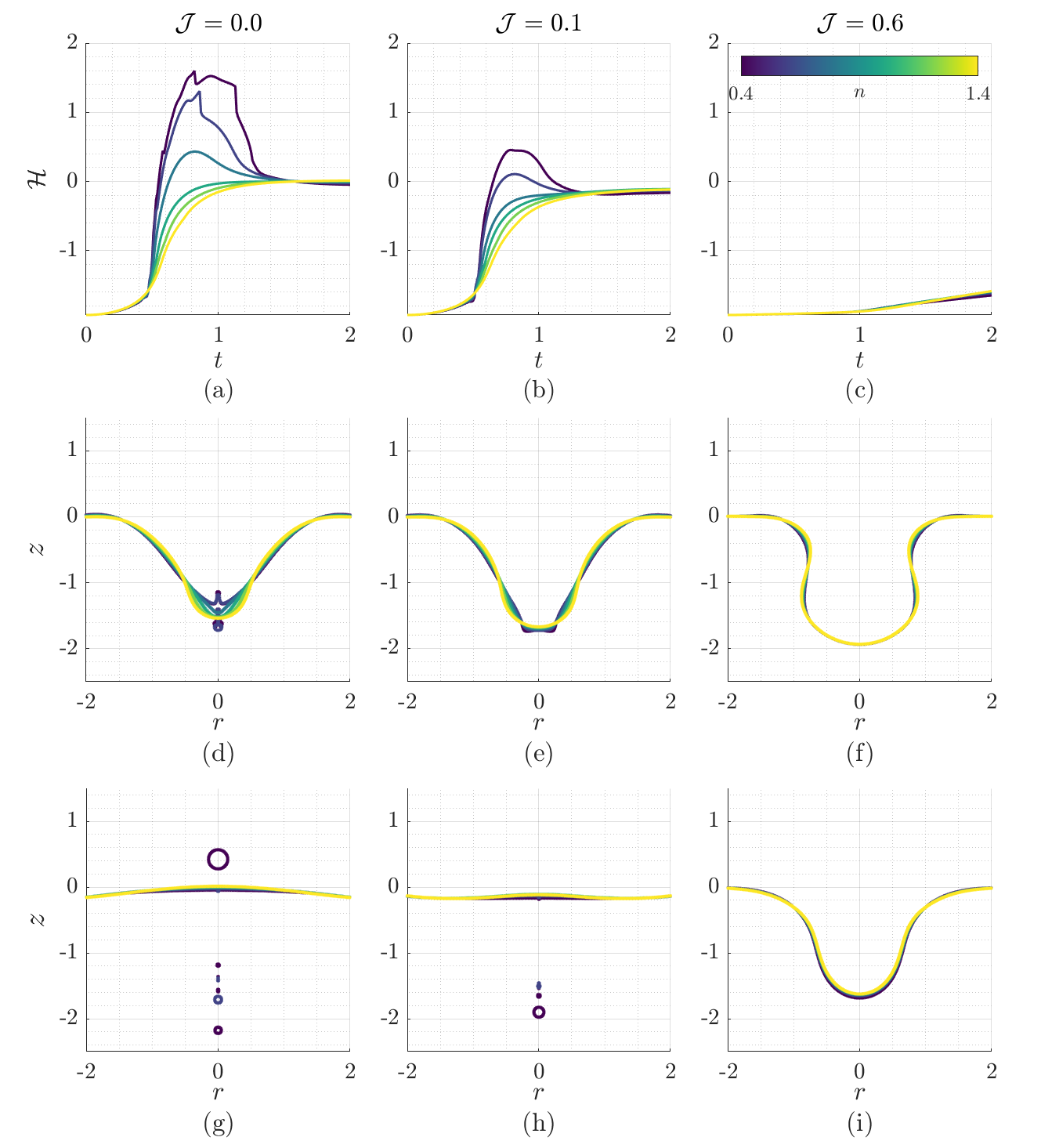}
\caption{Effect of flow behavior index $n$ for $\mathcal{J}=(0,0.1,0.6)$ with $Bo=0.01$ and $Oh_K=0.1$, corresponding to columns one-to-three, respectively. Panels (a–c) show the jet tip height $\mathcal{H}(t)$. Panels (d–f) and (g–i) display the interfacial shape at $t=0.5$ and $t=2.0$, respectively. The colour indicates the value of $n$, as specified in the legend in panel (c).}
    \label{fig:jet_height_n}
\end{figure}

\begin{figure}
    \centering
    \includegraphics[width=\linewidth]{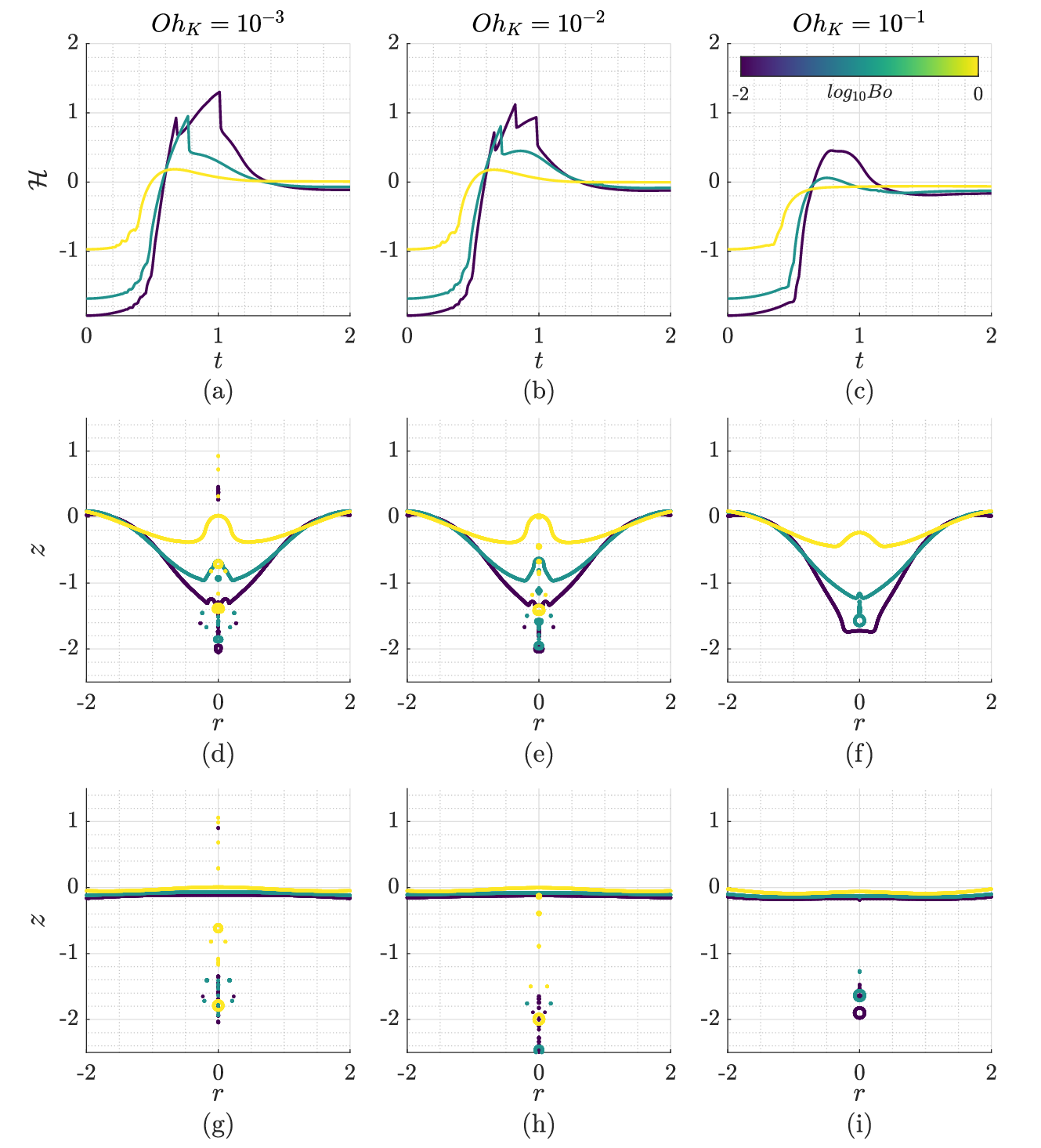}
    \caption{Effect of Bond number $Bo$ for $Oh=(0.001,0.01, 0.1)$ with $n=0.4$ and $\mathcal{J}=0.1$, corresponding to columns one-to-three, respectively. Panels (a–c) show the jet tip height $\mathcal{H}(t)$. Panels (d–f) and (g–i) display the interfacial shape at $t=0.5$ and $t=2.0$, respectively. The colour indicates the value of $Bo$, as specified in the legend in panel (c).}
    \label{fig:jet_height_bo}
\end{figure}

\begin{figure}
    \centering
    \includegraphics[width=\linewidth]{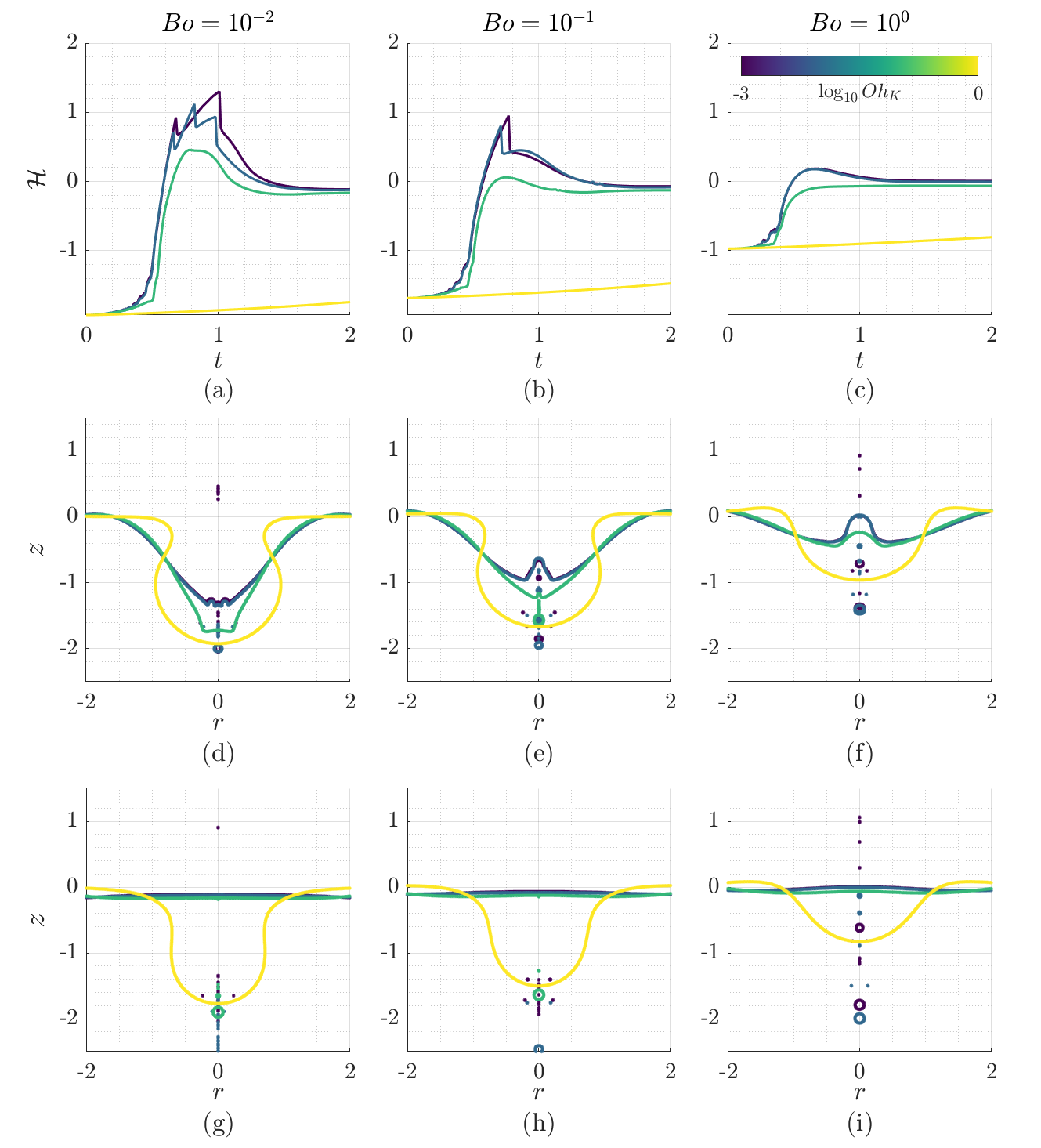}
    \caption{Effect of Ohnesorge number $Oh_K$ for $Bo=(0.01, 0.1, 1)$ with $n=0.4$ and $\mathcal{J}=0.1$, corresponding to columns one-to-three, respectively. Panels (a–c) show the jet tip height $\mathcal{H}(t)$. Panels (d–f) and (g–i) display the interfacial shape at $t=0.5$ and $t=2.0$, respectively. The colour indicates the value of $Oh_K$, as specified in the legend in panel (c).}
    \label{fig:jet_height_oh}
\end{figure}

Figure~\ref{fig:jet_height_shapes_J} highlights the influence of the plastocapillary number $\mathcal{J}$ on jet height and interfacial shape for a shear-thinning fluid ($n = 0.4$, left column), a Newtonian case ($n = 1.0$, middle column), and a shear-thickening fluid ($n = 1.4$, right column) with  $Bo=0.01$ and $Oh_K=0.1$.
For the shear-thinning case ($n=0.4$), the jet height decreases monotonically with increasing $\mathcal{J}$ as the growing yield stress progressively damps the collapse (see figure \ref{fig:jet_height_shapes_J}a).  At low $\mathcal{J}$, 
the cavity reversal produces a Worthington jet and subsequent droplet ejection, manifested as sharp oscillations in jet height. As   $\mathcal{J}$ increases, the maximum jet height decreases and eventually saturates, indicating that further increases in yield stress have little additional effect. The interfacial shapes  at $t = 0.5$ (figure~\ref{fig:jet_height_shapes_J}d) and $t = 2$ (figure~\ref{fig:jet_height_shapes_J}g) 
show the delay brought by the yield stress. 
%
The interfacial dynamics change dramatically depending on $\mathcal{J}$. Low $\mathcal{J}$ cases produce  jets with  with droplet detachment, whereas high $\mathcal{J}$ cases
remain nearly undeformed as the interface becomes rigid because yield stresses dominate the flow behaviour.
At $Oh=0.1$, the flow is predominantly viscous–capillary dominated.
In high-strain-rate regions such as the cavity bottom when its curvature reverses and the jet tip, the local shear rate becomes large, reducing the effective viscosity $\mu_{\mathrm{eff}}$ in shear-thinning fluids.
At large $\mathcal{J}$, the yield stress resists cavity collapse and imparts rigidity to the interface,  thereby  limiting the formation of high-shear-rate regions and weakening or preventing jet formation.
The competition between shear-thinning viscosity reduction (which promotes deformation) and the rigidification induced by yield stress behavior (which suppresses motion) governs the transition from droplet ejection at low $\mathcal{J}$ to minimal deformation at high $\mathcal{J}$ for $n = 0.4$.
For $n = 1.0$ (Bingham-like behavior) and $n = 1.4$ (shear-thickening), the collapse is strongly suppressed and no jetting or droplet formation occurs across the range of $\mathcal{J}$ considered, although the temporal interface evolution remains faster for $n = 1.0$ than for $n = 1.4$.  For $n > 1$, higher shear rates increase the effective viscosity of the fluid in regions of high deformation rate  slowing the motion, and enhance dissipation. Figure~\ref{fig:jet_height_shapes_J}f shows that the strong shear thickening has prevented the collapse, and subsequently the reverse of the cavity.
The results for $n=1$ are consistent with \citet{sanjay2021bursting}, who reported that the entire cavity  yields and the interface relaxes toward its initial level.
Overall, these observations demonstrate that shear-thinning fluids ($n<1$) favour jet formation at low $\mathcal{J}$ through local viscosity reduction, whereas Newtonian and shear-thickening fluids ($n\ge1$) remain dominated by yield and viscous resistance, suppressing jet dynamics.

Figure~\ref{fig:jet_height_n} shows how the rheological index $n$ influences the jet dynamics for three representative plastocapillary numbers, $\mathcal{J} = (0,\,0.1,\,0.6)$, with $Bo = 0.01$ and $Oh_K = 0.1$. 
For $\mathcal{J}=0$ (Newtonian limit), the system transitions across the three regimes as $n$ increases, from droplet detachment in shear-thinning fluids to complete suppression of jet formation in shear-thickening ones. For $\mathcal{J}=0.1$ (weakly yield-stress effects), only two regimes are observed as $n$ increases, with jet without droplets  giving way to arrested jetting. At $\mathcal{J}=0.6$, strong yield-stress effects dominate the stress balance, and the interfacial motion is almost entirely damped (figure~\ref{fig:jet_height_n}c), indicating that the yield stress prevents significant cavity collapse and jet acceleration. The corresponding jet heights also diminish systematically with increasing $\mathcal{J}$, reflecting the progressive damping of interfacial acceleration by the yield stress.
The interface profiles at $t=0.5$ and $t=2$ further illustrate how increasing $\mathcal{J}$ progressively slows the collapse and limits interfacial motion.
Even at early times, variations in $n$ markedly influence the collapse dynamics. Decreasing $n$ (shear-thinning) lowers the local viscosity in high-strain regions, promoting rapid flow focusing and the formation of tall, slender jets that may culminate in droplet detachment, whereas increasing $n$ (shear-thickening) enhances viscous resistance and damps jet growth. For $\mathcal{J}=0$ and $0.1$, this yields a smooth progression from  jetting at low $n$ to arrested motion at high $n$. At larger yield stresses, as shown in figures~\ref{fig:jet_height_n}f,i, the dynamics are almost completely suppressed, and the interfacial profiles become nearly independent of $n$.

Figure~\ref{fig:jet_height_bo} illustrates the influence of the Bond number  with $n=0.4$ and $\mathcal{J}=0.1$ for different values of $Oh_K$. 
As $Bo$ increases, the maximum jet height systematically decreases, consistent with the stronger role of gravity in flattening the cavity and opposing upward motion. As $Bo$ increases, the location of the bottom of the bubble shifts up due to the gravitational effects (compares figure~\ref{fig:jet_height_bo}a and \ref{fig:jet_height_bo}b).
This trend agrees with previous observations for Newtonian fluids, where larger $Bo$ reduces the cavity curvature and weakens the capillary focusing that drives the jet \citep{walls2015jet, berny2020role}. Moreover, as $Bo$ increases, the dynamics transition from droplet-producing jets to cases with no droplet ejection, similar to the suppression of jet drop ejection reported in Newtonian bursting at large $Bo$ \citep{berny2020role}. 
The bubble profiles at $t = 0.5$ and $t = 2$ further demonstrate these effects. At $t = 0.5$, low $Bo$ and low $Oh_K$ cases produce sharp, rapidly rising jets, whereas high $Bo$ and high $Oh_K$ cases remain broad and nearly undeformed. By $t = 2$, these differences persist, confirming that viscosity and gravity act together to suppress jet elongation, reduce droplet ejection, and slow the overall collapse dynamics.


Figure~\ref{fig:jet_height_oh} illustrates the  effect of the the Ohnesorge number $Oh_K$   for three representative Bo numbers, $Bo = (10^{-2},10^{-1},10^{0})$, with $n=0.4$ and $\mathcal{J}=0.1$.
Increasing  $Oh_K$ systematically suppresses the cavity collapse by enhancing viscous dissipation.
At $Oh_K = 0.001$ (nearly inviscid), viscous dissipation is negligible, allowing capillary waves generated during the collapse to focus efficiently at the axis. 
This strong momentum   produces faster jets whose tip  pinch-off to result in  a smaller number of  droplets (figure~\ref{fig:jet_height_oh}a). 
At $Oh_K = 0.01$,  partial damping of these waves reduces axial focusing, yielding thinner, more elongated jets and a higher frequency of droplet detachment (figure~\ref{fig:jet_height_oh}b). 
Further increasing $Oh_K$ enhances viscous dissipation throughout the bulk, weakening the collapse and nearly suppressing jet formation (figure~\ref{fig:jet_height_oh}c). 
The corresponding bubble profiles at $t = 0.5$ and $t = 2$ further illustrate these effects. 

\newcommand{\OhK}{\mathrm{Oh}_{K}}
\newcommand{\Bo}{\mathrm{Bo}}

\begin{figure}
    \centering
\begin{tabular}{cc}
\includegraphics[width=0.5\linewidth]{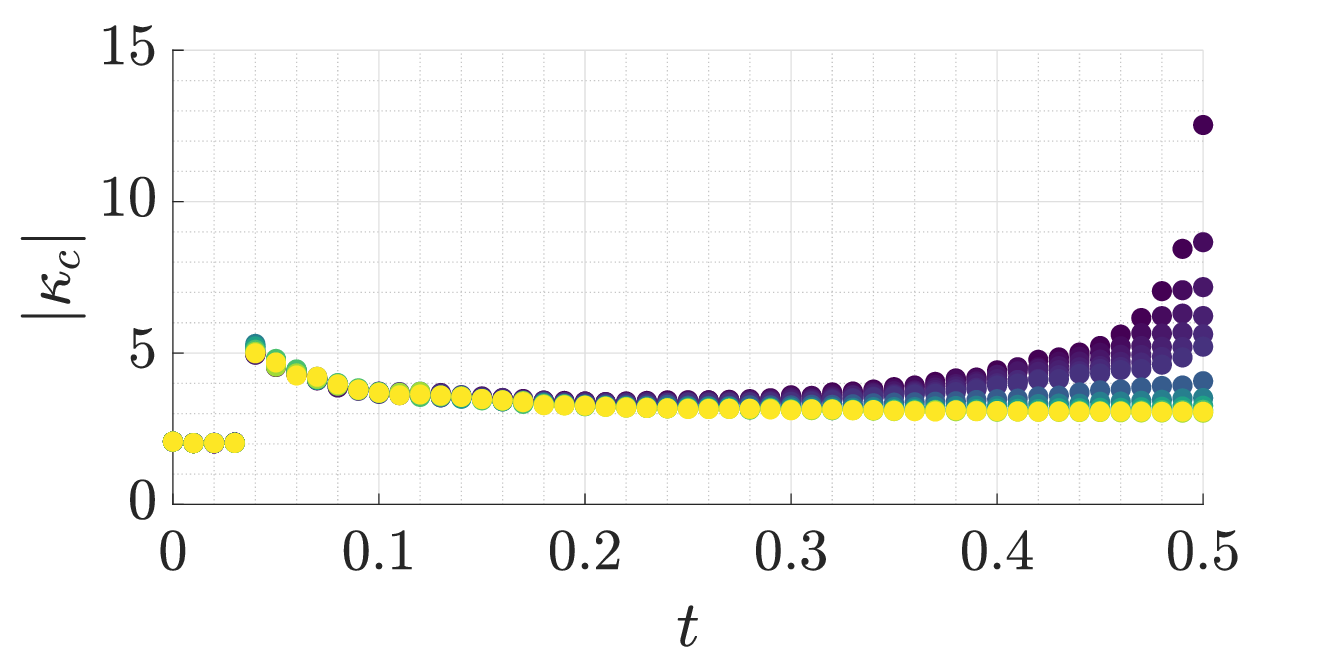} &
\includegraphics[width=0.5\linewidth]{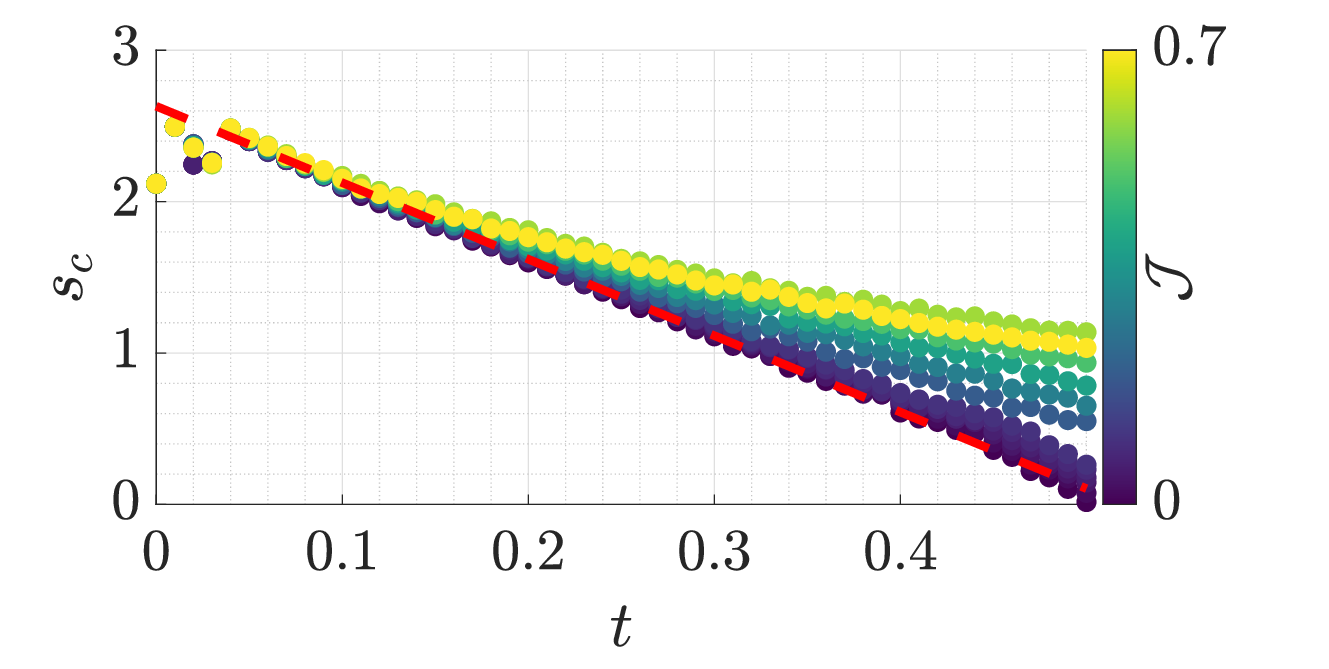}\\
(a) & (b) \\
\\[-0.5em]
\includegraphics[width=0.5\linewidth]{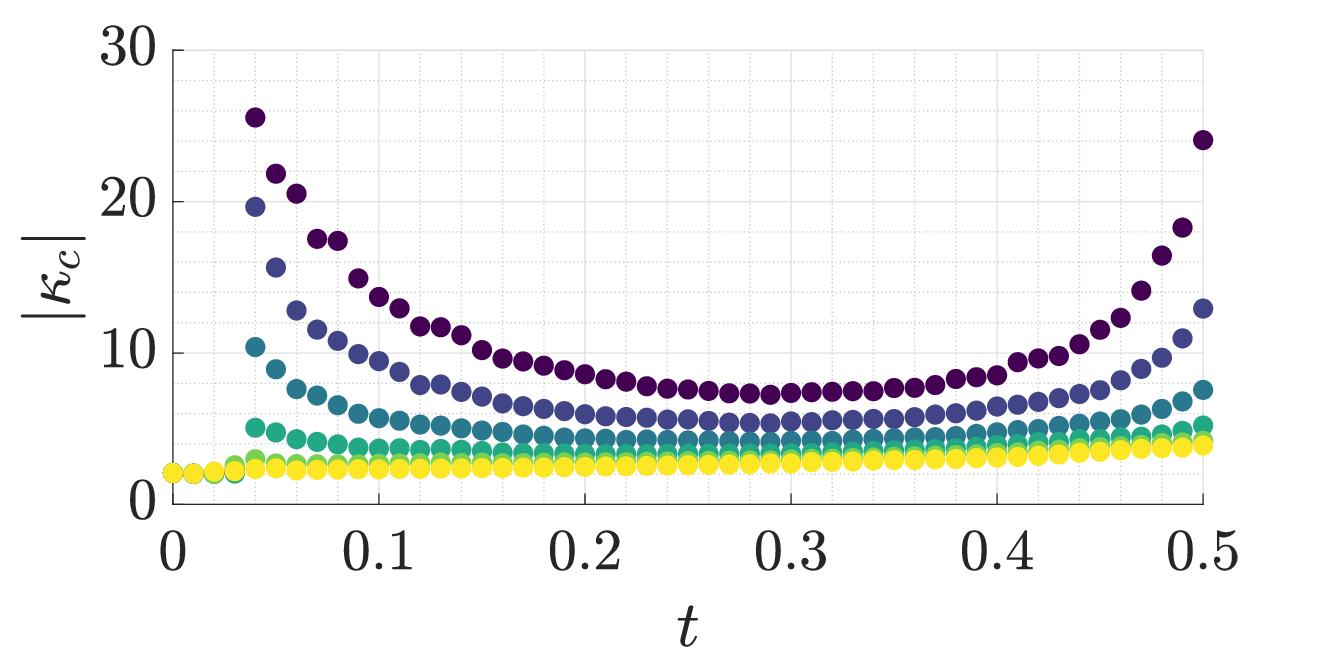} &
\includegraphics[width=0.5\linewidth]{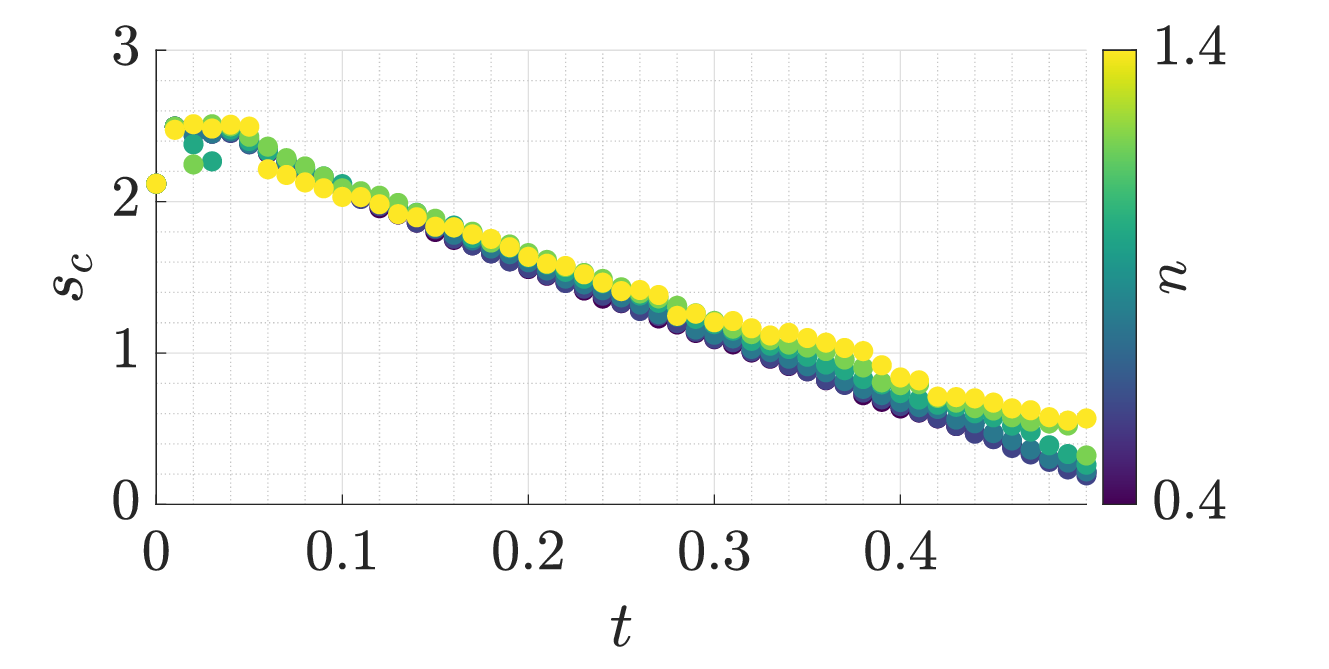}\\
(c) & (d) \\
\\[-0.5em]
\includegraphics[width=0.5\linewidth]{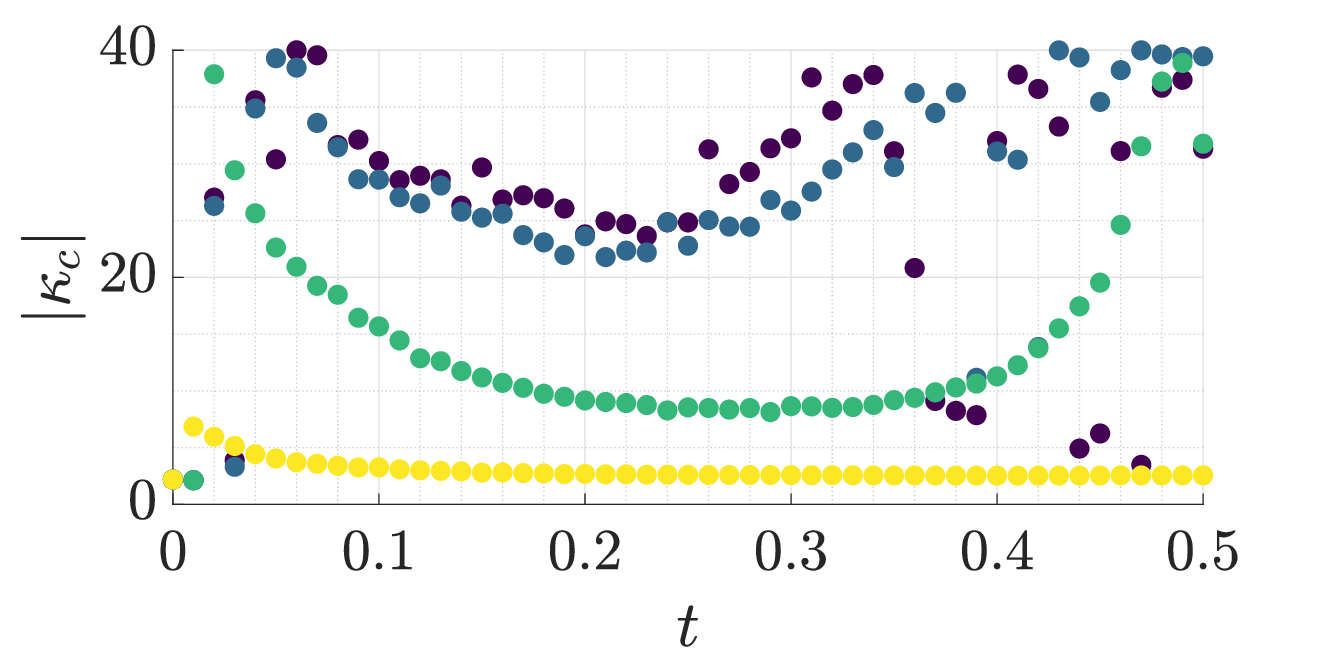} &
\includegraphics[width=0.5\linewidth]{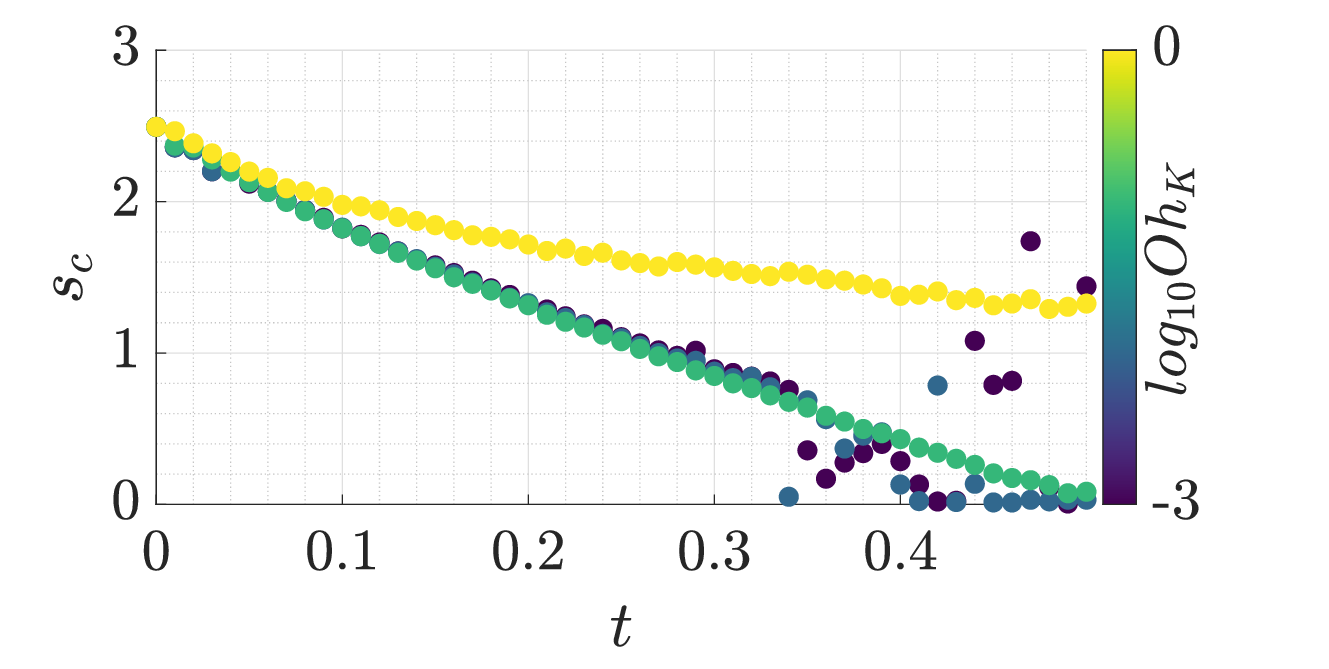}\\
(e) & (f) \\
\\[-0.5em]
\includegraphics[width=0.5\linewidth]{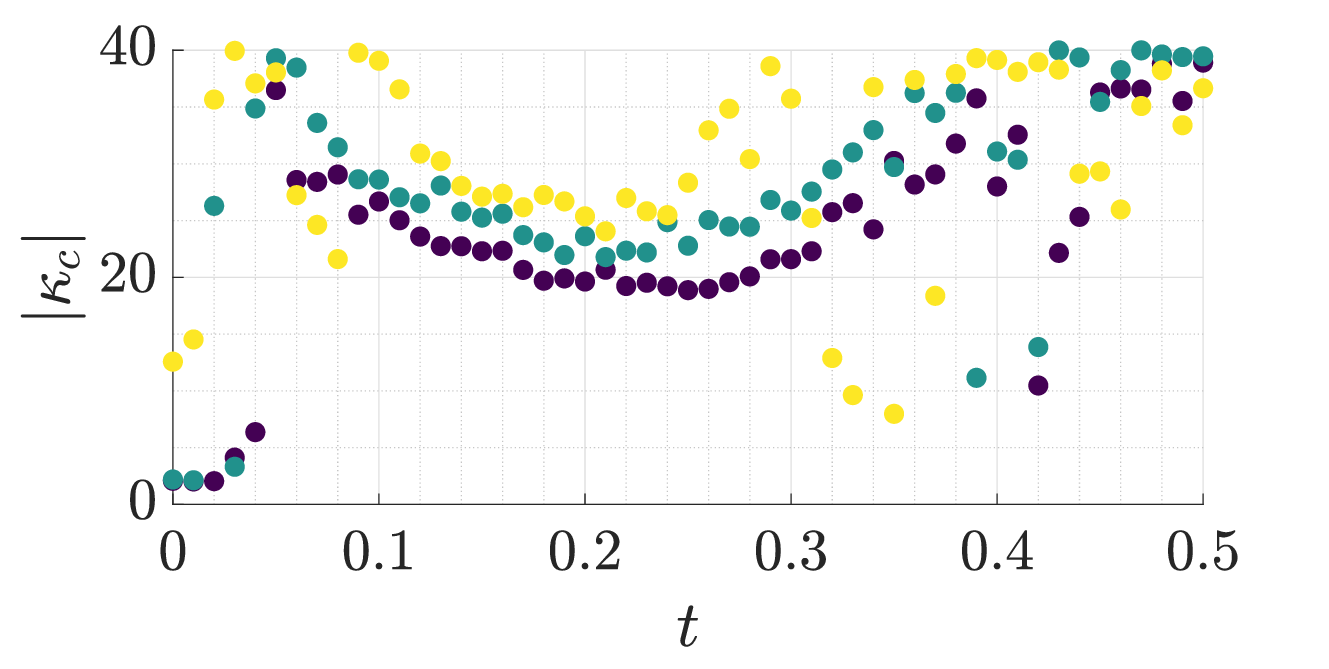} &
\includegraphics[width=0.5\linewidth]{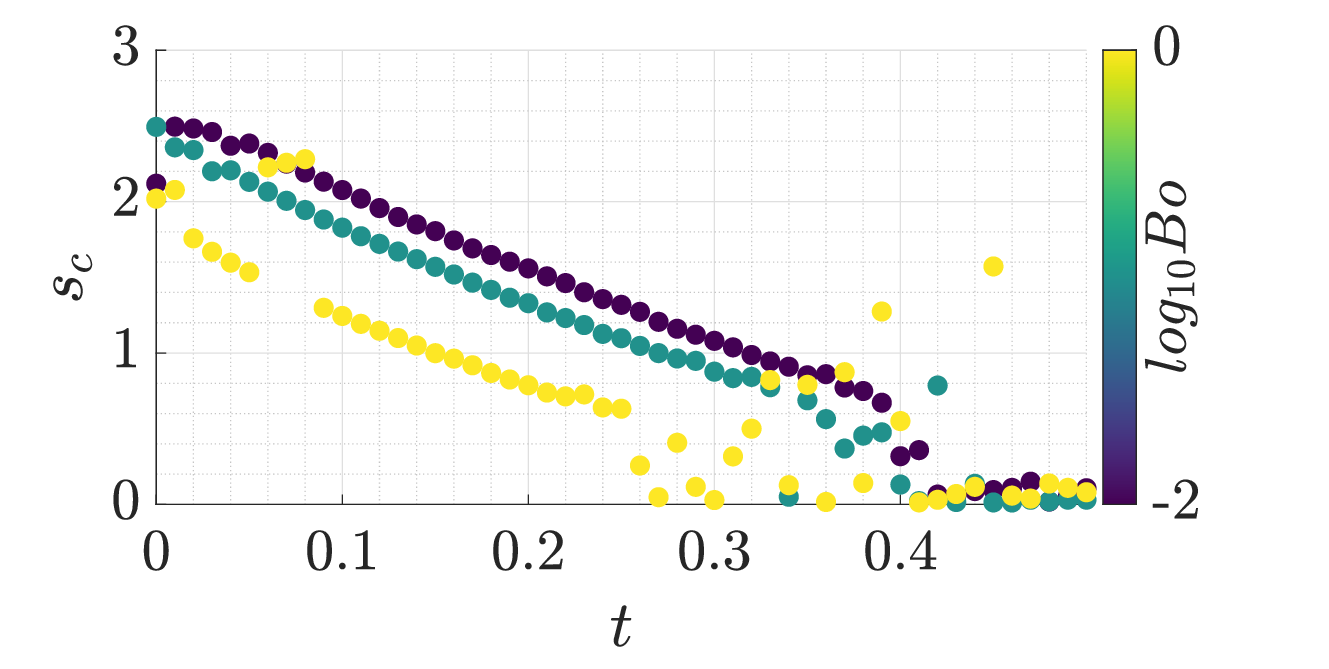}\\
(g) & (h) \\
\end{tabular}

\caption{Temporal evolution of capillary waves illustrating the effect of different governing parameters.
In each panel, the left plot shows the curvature magnitude $|\kappa_c|$ and the right plot shows the crest arclength $s_c$ as functions of time.
Panels (a,b) vary the plastocapillary number $\mathcal{J}$ ($n=1.0$, $\Bo=0.01$, $\OhK=0.1$);
(c,d) vary the flow behavior index $n$ ($\mathcal{J}=0.1$, $\Bo=0.01$, $\OhK=0.1$);
(e,f) vary the Ohnesorge number $\OhK$ ($n=0.4$, $\mathcal{J}=0.1$, $\Bo=0.1$);
and (g,h) vary the Bond number $\Bo$ ($n=0.4$, $\mathcal{J}=0.1$, $\OhK=0.01$).
}
\label{fig:capillary_waves_all}
\end{figure}

Next, we examine the evolution of the capillary waves generated during the bursting process, which play a central role in the characteristics of the jet formed  \citep{gordillo2019capillary}. During the cavity collapse, the interplay between viscous and capillary effects selects a dominant wave that will control the jet formation and  its velocity. In this work, we follow \citet{gordillo2019capillary} and \citet{sanjay2021bursting} who   track the strongest wave by chasing the maximum curvature of the interface $|\kappa_c|$. The location of this wave on the cavity is denoted by the angular position $s_c$, which
is measured from the cavity bottom apex to the wave crest location. 

Figure~\ref{fig:capillary_waves_all}a,b shows the effect of the plastocapillary number $\mathcal{J}$ (with fixed $n=1.0$, $Bo=0.01$, and $Oh=0.1$). 
In all cases, the magnitude of the crest curvature, $|\kappa_c|$, initially decreases to a similar minimum value as the sharp rim relaxes, followed by a rise associated with the formation of the capillary wave (see figure \ref{fig:capillary_waves_all}a). 
For small $\mathcal{J}$, this increase is sharp and pronounced, whereas for larger  $\mathcal{J}$, it progressively flattens as more of the fluid remains unyielded, and the available energy is expended in yielding rather than accelerating the interface.
The arclength $s_c$ decreases monotonically in all cases (see figure \ref{fig:capillary_waves_all}b), reflecting the inward motion of the crest, but the magnitude of $s_c$ diminishes
as $\mathcal{J}$ increases,
consistent with the progressive rigidifcation of the interface due to higher yield stress (in agreement with the previous work by \citet{sanjay2021bursting} in Bingham fluids). 
The red dashed line in figure~\ref{fig:capillary_waves_all}b corresponds to the Newtonian limit ($\mathcal{J}\to0$), illustrating the asymptotic behavior recovered as the yield stress vanishes.

Figure~\ref{fig:capillary_waves_all}c,d presents the effect of the flow behavior index $n$ (for $\mathcal{J}=0.1$, $Bo=0.01$, and $Oh_K=0.1$). Shear thinning cases lead to larger values of $|\kappa_c|$ throughout the wave propagation process, reflecting stronger capillary focusing. This arises because the wave region experiences high shear rates, which locally reduce the viscosity and thus weaken viscous damping, allowing sharper interfacial deformation
In contrast, 
shear thickening cases exhibit smaller maximum curvatures due to enhanced viscous resistance. The arc-length $s_c$ exhibits only modest changes with $n$, showing that the shear-thinning/thickening effect mainly affects the focusing strength rather than the lateral propagation of the wave.

Finally, panels (e,f) depict the influence of the Ohnesorge number $Oh_K$ for $Bo=0.01$, $n=0.4$, and $\mathcal{J}=0.1$.
As $Oh_K$ decreases, the curvature $|\kappa_c|$ grows more rapidly and attains larger peaks, indicative of stronger capillary focusing and reduced viscous dissipation.
Conversely, at large $Oh_K$, viscous effects dominate, leading to a weak curvature response and a gradual relaxation of the interface.
The arclength $s_c$ decreases monotonically in all cases, reflecting the inward motion of the crest; however, the two lowest $Oh_K$ cases nearly coincide, suggesting that once viscous damping becomes negligible, the dynamics asymptote to an inertia–capillary regime where viscosity no longer alters the wave propagation.

Finally, figure~\ref{fig:capillary_waves_all}g,h illustrates the effect of the Bond number $Bo$ for $Bo=0.1$, $n=0.4$, and $\mathcal{J}=0.1$.
Increasing $Bo$ raises the overall level of $|\kappa_c|$ without altering its qualitative evolution, indicating that gravity sets the baseline curvature of the cavity while leaving the capillary–yield focusing mechanism largely unchanged. The corresponding $s_c$ curves shift downward and leftward with increasing $Bo$, consistent with a shallower cavity and shorter focusing distance due to gravitational flattening. Although $Bo$ controls the geometric scale of deformation, its influence couples weakly with the yield-stress and viscous effects established in the previous cases, larger $Bo$ slightly suppresses curvature growth at fixed $\mathcal{J}$ and enhances the viscous retardation of crest motion. Hence, in this regime, gravity chiefly modulates the extent of focusing, whereas yield stress and viscosity continue to dictate its strength and timescale.


\section{Conclusions \label{Con}}

We have investigated bubble bursting in Herschel–Bulkley media using fully resolved numerical simulations performed with the Basilisk framework, complemented by laboratory experiments with Carbopol gels that serve to validate the numerical predictions.
These numerical simulations  capture the entire sequence of events, from cavity collapse to jet formation, revealing how yield stress, shear-dependent rheology, viscosity, and gravity collectively reorganize capillary wave focusing and the ensuing dynamics.

A systematic exploration of the $(\mathcal{J}, n, Oh_K, Bo)$  parameter space with over 800 simulations exposes clear and interpretable trends. The plastocapillary number $\mathcal{J}$  promotes interfacial yielding, cavity collapse, and jetting, where large $\mathcal{J}$ leads to arrested interfaces and non-flat equilibria.
The flow index $n$ modulates this control by altering the apparent viscosity in high strain regions. Shear thinning with $n<1$ lowers the local resistance near the crest and rim, enabling tall and slender jets at small $\mathcal{J}$, whereas for intermediate Ohnesorge numbers, $Oh_K \approx 10^{-1}$, shear-thickening ($n>1$) or Newtonian behavior increases dissipation and suppresses jet formation. At very low $Oh_K$, however, inertia dominates and jets may still emerge even for shear-thickening fluids


Analysis of capillary waves provides a better understanding  of these trends.
With stronger yield stress and greater shear thickening effects,  the maximum curvature increases and the subsequent wave focusing becomes weaker. The bubble bursting therefore transitions from a Newtonian-like bursting which produces a vigorous jet, to a viscoplastic arrest which suppresses jet formation or even forms an unyielded cavity. 
The curvature relaxation  and subsequent focusing are progressively damped by yield stress and shear thickening, reflecting the transition from Newtonian-like bursting to viscoplastic arrest.
The arclength from the cavity bottom to the crest decreases in time for all cases, indicating inward migration of the wavefront. The speed of wave propagation is largest at small $\mathcal{J}$ and smallest at large $\mathcal{J}$, consistent with progressive arrest of motion. Gravity shifts the baseline curvature and shortens the geometric focusing distance as $Bo$ increases, which reduces jet height and the likelihood of droplet formation without changing the qualitative sequence of wave evolution.


The present study has several limitations. First, the thin cap is removed as part of the initial bubble shape, isolating the post-rupture dynamics. This assumption is valid when the Bond number is small, but we acknowledge that cap retraction at large $Bo$  can influence the bursting sequence. 
Second, the simulations are axisymmetric and neglect viscoelastic effects, whereas the experiments reveal weak elasticity.
Future work should incorporate the cap rupture process, three-dimensional simulations that capture azimuthal asymmetries, and Herschel–Bulkley fluids with controlled elasticity and surfactant effects. These extensions will refine the regime boundaries and provide quantitative criteria for jet formation and droplet statistics in complex non-Newtonian environments.

\medskip
\noindent\textbf{Declaration of Interests.} The authors report no conflict of interest.

\subsection*{Acknowledgments }
This research used the Delta advanced computing and data resource which is supported by the National Science Foundation (award OAC 2005572) and the State of Illinois. Z.Y. and J.F. acknowledge partial support by the NSF under grant no. CBET 2426809. Z.Y. and J.F. thank helpful discussions and assistance for rheological measurement from Mohammed Tanver Hossain at University of Illinois Urbana-Champaign.

\section*{Appendix: Validation of the Herschel--Bulkley formulation and sensitivity to the regularization parameter $\varepsilon$}

\begin{figure}
    \centering
    \includegraphics[width=0.8\linewidth]{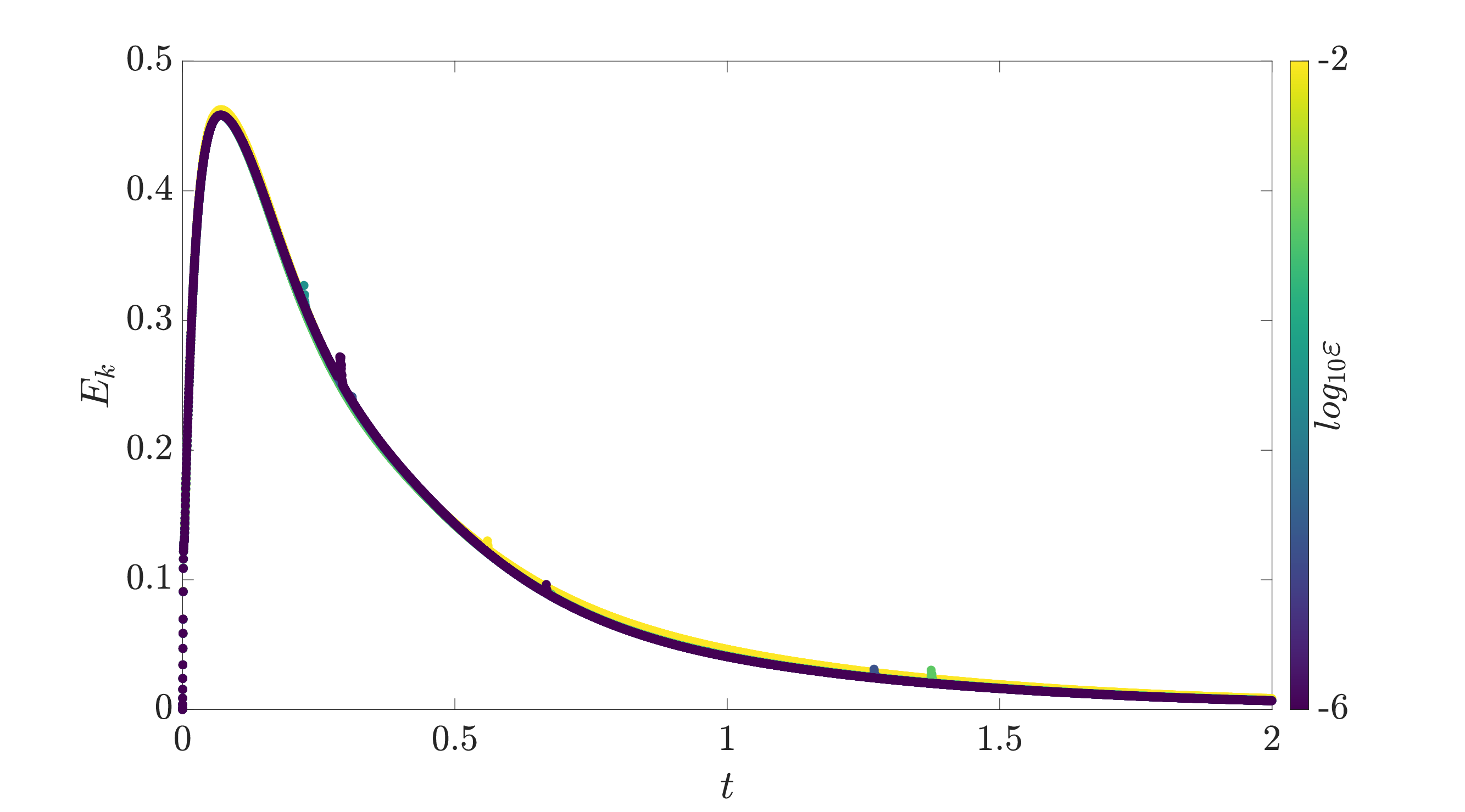}
    \caption{Sensitivity of the regularised Herschel--Bulkley model to the regularization parameter $\varepsilon$, evaluated through the temporal evolution of the kinetic energy for
    $n=0.4187$, $\mathcal{J}=0.1053$, $Bo=1.0990$ and $Oh_K=0.7015$ at $\varepsilon=10^{-2}-10^{-6}$.}
    \label{fig:epsilon_sensitivity}
\end{figure}

In the present work a regularized form of the Herschel–Bulkley model was adopted to avoid singularities in the apparent viscosity as the strain rate magnitude $\|\mathbf{\mathcal{D}}\|\to 0$.  The regularization parameter $\varepsilon$ controls the transition between yielded and unyielded regions. To assess the sensitivity of the results to $\varepsilon$, we performed a series of simulations for representative base cases using $\varepsilon = 10^{-2}$, $10^{-3}$, $10^{-4}$, $10^{-5}$, and $10^{-6}$. Figure~\ref{fig:epsilon_sensitivity} shows the evolution of the kinetic energy, defined as $E_k = \frac{1}{2} \int_V \| \mathbf{u} \|^2 \, dV
$, as a function of time and $\varepsilon$ for the experimental case from figure 2  (e.g., $n = 0.4187$, $J=0.1053$, $Bo=1.0990$ and $Oh_K=0.7015$). The results show that the regularized model reaches 
convergence for $\varepsilon \le 10^{-2}$, as 
further reduction of $\varepsilon$
produces no appreciable physical differences.
Based on these tests, we select $\varepsilon = 10^{-2}$ as the optimal compromise between numerical stability and accurate resolution of the yield surface. This choice ensures that the simulated regime maps and flow fields remain physically consistent while maintaining computational efficiency across the full parameter space explored.

For further validation of  the Herschel–Bulkley implementation, we direct the reader to \citet{yang2025pinch}, who investigated the breakup of a Newtonian thread surrounded by a viscoplastic fluid. They reported the temporal evolution of the neck radius for both the Newtonian case and the Herschel–Bulkley model with $\mathcal{J}=0$ and $n=1$. The close agreement between these two cases, including the recovery of the characteristic Newtonian scaling laws, confirms the accuracy of the Herschel–Bulkley formulation adopted in the present study. Moreover, \citet{yang2025pinch} demonstrated that the numerical method successfully reproduces the scaling behaviour of both the neck radius and the maximum axial velocity for shear-thinning and shear-thickening fluids.

\bibliographystyle{jfm}
\bibliography{jfm}

\end{document}